\documentclass[sigplan,screen]{acmart}

\AtBeginDocument{%
  }

\usepackage{algorithm}
\usepackage{algpseudocode}

\usepackage{tikz}
\usepackage{amsmath}
\usepackage{subfigure}
\usepackage{multirow} 
\usepackage{xspace} 
\usepackage{tcolorbox}
\usepackage{enumitem}

\newcommand{\method}[0]{{\textsc{ClawMobile}}\xspace}

\begin{document}

\copyrightyear{2026}
\acmYear{2026}
\setcopyright{cc}
\setcctype{by}
\acmConference[EuroMLSys '26]{Sixth European Workshop on Machine Learning and Systems }{April 27--30, 2026}{Edinburgh, Scotland Uk}
\acmBooktitle{Sixth European Workshop on Machine Learning and Systems (EuroMLSys '26), April 27--30, 2026, Edinburgh, Scotland Uk}
\acmDOI{10.1145/3805621.3807655}
\acmISBN{979-8-4007-2605-7/2026/04}

\title[\method: Rethinking Smartphone-Native Agentic Systems]{\method: Rethinking Smartphone-Native \\ Agentic Systems}


\author{Hongchao Du}
\authornote{These authors contributed equally to this work.}
\affiliation{%
  \institution{MBZUAI}
  \city{Abu Dhabi}
  \country{United Arab Emirates}
}

\author{Shangyu Wu}
\authornotemark[1]
\affiliation{%
  \institution{MBZUAI}
  \city{Abu Dhabi}
  \country{United Arab Emirates}
}

\author{Qiao Li}
\affiliation{
  \institution{MBZUAI}
  \city{Abu Dhabi}
  \country{United Arab Emirates}
}

\author{Riwei Pan}
\affiliation{
  \institution{City University of Hong Kong}
  \city{Hong Kong}
  \country{China}
}

\author{Jinheng Li}
\affiliation{
  \institution{Independent Researcher}
  \city{Hong Kong}
  \country{China}
}

\author{Youcheng Sun}
\affiliation{%
  \institution{MBZUAI}
  \city{Abu Dhabi}
  \country{United Arab Emirates}
}

\author{Chun Jason Xue}
\affiliation{%
  \institution{MBZUAI}
  \city{Abu Dhabi}
  \country{United Arab Emirates}
}
\renewcommand{\shortauthors}{Du et al.}

\begin{abstract}
Smartphones represent a uniquely challenging environment for agentic systems. Unlike cloud or desktop settings, mobile devices combine constrained execution contexts, fragmented control interfaces, and rapidly changing application states. As large language models (LLMs) evolve from conversational assistants to action-oriented agents, achieving reliable smartphone-native autonomy requires rethinking how reasoning and control are composed.

We introduce \method as a concrete exploration of this design space. \method adopts a hierarchical architecture that separates high-level language reasoning from structured, deterministic control pathways, improving execution stability and reproducibility on real devices. Using \method as a case study, we distill the design principles for mobile LLM runtimes and identify key challenges in efficiency, adaptability, and stability. We argue that building robust smartphone-native agentic systems demands principled coordination between probabilistic planning and deterministic system interfaces. The implementation is open-sourced~\footnote{\url{https://github.com/ClawMobile/ClawMobile}} to facilitate future exploration.

\end{abstract}





\maketitle
\section{Introduction}

Smartphones are emerging as a new frontier for agentic systems~\cite{liu2025llmpowered, nguyen-etal-2025-gui,wang2024mobile, appagent}. As large language models (LLMs) transition from conversational assistants to action-oriented agents, enabling reliable autonomy on mobile devices becomes both an opportunity and a system challenge~\cite{ai_agent_lat, fan2025core, zhou2025mai}.

Recent work on mobile agents spans two overlapping dimensions: how agents act (UI-based interaction versus tool/API-based control) and where they run (cloud-hosted versus on-device)~\cite{wang2024mobile, openclawandroid, appagent, liu2024autoglm}. Many systems pair LLM planning with UI automation to operate across arbitrary apps~\cite{gou2024navigating, zheng2024gpt, hoscilowicz2025clickagent, wang2025ponder}, while others emphasize on-device execution for latency, privacy, and continuous assistance~\cite{wu2024first, jiang2025lightagent}. 
This overlap reveals a practical tension: UI interaction offers broad coverage but is sensitive to UI drift and timing, whereas tool/API control is more stable and verifiable but incomplete across apps and device functions. 
As a result, real-world mobile autonomy often fails not because the agent cannot plan, but because execution is repeatedly disrupted by variable execution conditions of the mobile devices.

In this work, we present \method, which addresses the action-side of this design space with a hierarchical runtime architecture: a high-level LLM orchestrator decomposes goals and coordinates lower-level agents (including UI agents) alongside tool-based, deterministic device control backends. 
This design targets real-world reliability issues on phones by coordinating tool-based control and UI interaction with explicit progress verification and recovery from common mobile interruptions such as permission prompts, backgrounding, and transient UI changes.
Experimental results on six real-life tasks show that \method achieves near-perfect task completion performance while maintaining acceptable efficiency.
Finally, we comprehensively discuss the key research questions of achieving mobile autonomy.

In summary, this paper makes the following contributions:
\begin{itemize}[topsep=0pt,itemsep=-1ex,partopsep=1ex,parsep=1ex]
\item We present a hierarchical architecture where a high-level agent coordinates lower-level agents and tool-based device controls for mobile task execution.
\item We instantiate this architecture in \method and report preliminary results and observations of real-device deployments.
\item We identify key challenges in smartphone-native agents and outline a research agenda spanning efficiency, adaptability, and stability.
\end{itemize}

\section{Background}

\subsection{UI-Centric Mobile Agent Automation}

Recent progress in multimodal LLM agents has renewed interest in mobile autonomy, i.e., agents capable of executing multi-step tasks within the rich app ecosystems on smartphones. A growing body of work applies LLM-based interaction planning and function-calling capabilities to smartphone environments, where agents must interpret interface states and invoke corresponding actions to satisfy user intent~\cite{liu2025llmpowered, nguyen-etal-2025-gui}. Systems such as Mobile-Agent~\cite{wang2024mobile} and AppAgent~\cite{appagent} demonstrate multimodal reasoning for mobile GUI interaction, emphasizing visual grounding and structured action generation over app interfaces. These works illustrate the feasibility of LLM-driven control over smartphone applications, but also highlight the difficulty of generalizing across heterogeneous UI layouts and evolving app ecosystems.

A complementary line of work focuses on end-to-end mobile automation powered by LLMs, shifting attention from interaction primitives to task-level completion. In this setting, LLM planners are tightly coupled with UI-facing control layers to execute realistic, multi-step workflows. This perspective has motivated the development of standardized evaluation environments such as AndroidWorld~\cite{rawles2025androidworld} and SPA-Bench~\cite{ICLR2025_9a75f4dd}, which assess agent performance under task-driven conditions. Systems like AutoDroid~\cite{AutoDroid, wen2023empowering} further integrate automated exploration and app-specific knowledge to improve task completion rates. 
Practical frameworks such as DroidRun~\cite{droidrun} similarly integrate LLMs' planning with accessibility and device interfaces, and support synthesizing executable code for multi-step operations rather than relying solely on step-wise UI actions. 
While these approaches demonstrate improved coverage and benchmarking rigor, they remain primarily UI-centric and continue to rely heavily on probabilistic reasoning over dynamic interface states.

In summary, this body of work demonstrates the potential of LLM-driven mobile interaction while revealing persistent limitations in robustness, efficiency, and runtime control.

\subsection{OpenClaw Agent Framework}

OpenClaw~\cite{openclaw} is an open-source, self-hosted agent framework that connects large language models to external tools and messaging surfaces through a unified runtime. Originally developed as a personal AI assistant, OpenClaw has rapidly grown into a general-purpose agent platform since its launch in late 2025.

At its core, OpenClaw operates through a central Gateway process that multiplexes WebSocket and HTTP communication on a single port. The Gateway manages session lifecycles, tool dispatch, channel routing, and agent orchestration. Agents execute through an embedded runtime that implements a multi-turn reasoning loop: the model receives conversation history and system prompts, generates tool calls, executes them sequentially, and resubmits updated context until a text-only response terminates the loop. Tool invocation follows a multi-level policy cascade (global, provider, agent, session, sandbox), enabling fine-grained permission control over execution.

A distinctive architectural feature is OpenClaw's node system. Device-specific nodes, including macOS, iOS, and Android clients, connect to the Gateway via WebSocket and expose local capabilities (e.g., camera capture, screen recording, system notifications) as remotely invocable tools. This design enables cross-device agent workflows while maintaining a centralized reasoning and control plane.
Several community-driven efforts have further extended OpenClaw to mobile devices~\cite{clawphone, openclawandroid, phoneclaw}, covering a spectrum of deployment strategies: remote orchestration through a companion node that delegates reasoning to a cloud-hosted Gateway~\cite{openclawandroid}, full on-device hosting via Termux with local sensor access~\cite{clawphone}, and UI-centric automation using the Android Accessibility Service~\cite{phoneclaw}. These extensions demonstrate the versatility of OpenClaw's architecture on mobile platforms, yet none systematically unifies deterministic system control with probabilistic UI reasoning within a single mobile runtime. This gap motivates the design of \method, which treats such coordination as a first-class system problem.

\begin{figure}[b]
  \centering
  \includegraphics[]{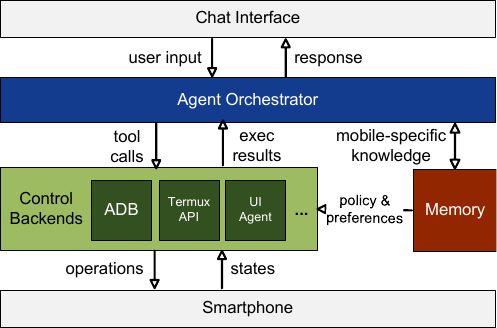}
  \caption{~\method~architecture. The Agent Orchestrator serves as the central coordination layer. Control Backends provide structured execution interfaces to the smartphone. Memory maintains mobile-specific knowledge and execution preferences that guide runtime behavior.}
  \label{fig:architecture}
  \Description{\method~architecture.}
\end{figure}

\section{\method}
This section introduces the design of \method, which is a smartphone-native agent runtime that treats the phone as the primary execution environment. 

\subsection{System Architecture}

Figure~\ref{fig:architecture} illustrates the overall architecture of \method.
\method is designed to separate high-level language reasoning from low-level device execution. Rather than embedding all interaction logic within a monolithic agent loop, \method decomposes mobile autonomy into three cooperating components:
\begin{itemize}
\item \textbf{Agent Orchestrator}: the top-level reasoning loop that converts user input into an executable plan.
\item \textbf{Control Backends}: pluggable execution modules that expose action primitives with defined semantics.
\item \textbf{Memory}: runtime knowledge that biases backend choice and encodes mobile-specific operating steps.
\end{itemize}

At the top of the stack, the Chat Interface serves as the entry point for high-level tasks.
The orchestrator will interpret user input, generate task-level plans, and coordinate subsequent execution steps.
The Agent Orchestrator does not directly manipulate device interfaces. Instead, it interacts with a set of Control Backends through explicit tool calls. These backends encapsulate concrete execution mechanisms, such as structured system commands (ADB), hardware-level APIs (Termux API~\cite{termuxapi}), or semantic UI interaction modules. Each backend exposes well-defined action primitives and returns bounded execution results. By routing execution through these backends, the orchestrator maintains a clear separation between reasoning and actuation.

Control Backends interact with the Smartphone by issuing operations and receiving state updates. The architecture explicitly models this bidirectional relationship: backends produce operations that affect device state, and in turn receive structured state signals that inform subsequent reasoning steps. This explicit state loop prevents the orchestrator from inferring execution success purely through LLM reasoning.

In parallel, a Memory component provides mobile-specific knowledge, execution preferences, and policy signals to the orchestrator. Rather than being tightly coupled to any single backend, the memory component acts as an auxiliary runtime layer that influences backend selection and task decomposition. This design enables domain specialization and adaptive behavior without entangling long-term knowledge with low-level control logic.

Taken together, the architecture separates intent interpretation, device interaction, and memory management into distinct components. This decomposition enables the modular extension of control pathways, explicit execution boundaries, and clearer reasoning about runtime behavior in heterogeneous mobile environments.

\subsection{Control Backends}

Smartphone-native autonomy cannot rely on a single control surface. \method treats smartphone control as inherently heterogeneous. Mobile devices expose both deterministic interfaces (system-level commands, structured APIs) and semi-structured UI surfaces that require semantic interpretation. \method supports multiple classes of control backends, each corresponding to a distinct execution paradigm within smartphone-native agents. Rather than treating all device interactions uniformly, the runtime differentiates between deterministic system interfaces, semantic UI agents, and direct UI control actions. These backends provide complementary capabilities with different trade-offs in predictability, coverage, and execution cost.

\textbf{Deterministic Backends.} Deterministic backends (e.g., ADB and Termux API~\cite{termuxapi}) expose structured system interfaces with well-defined semantics. These include operations that can be executed through stable device-level commands or hardware APIs. Because their outcomes are predictable and bounded, deterministic backends provide strong execution guarantees and minimal reasoning overhead. When a task can be satisfied through such structured primitives, the runtime prioritizes this pathway.

\textbf{UI Agents.} UI agents operate over visually semantic representations of the current screen state. By interpreting interface structure and textual content, they enable flexible interaction across heterogeneous applications. However, this flexibility comes at the cost of higher uncertainty and increased context requirements. UI agents are therefore invoked selectively when deterministic pathways cannot fully satisfy the requested task.

\textbf{Direct UI Control Actions.} In cases where neither structured backends nor semantic UI agents provide sufficient coverage, the runtime may resort to lower-level UI control actions. These actions operate directly on interface elements without extensive reasoning. While broadly applicable, they offer weaker guarantees regarding task-level semantics and are treated as fallback mechanisms.

By distinguishing these backend categories, \method makes execution modality explicit. This separation lays the foundation for runtime policies that reason about backend selection, cost, and reliability, rather than embedding such decisions implicitly within a monolithic agent loop.

\begin{table*}[t]
\centering
\caption{Completion ratios and end-to-end execution time on six real-life tasks (DR: DroidRun; CM-w/o-DR: \method without DroidRun; CM: \method; Timeout = 600s).}
\label{tab:main}
\begin{tabular}{lp{7cm}cccccc}
\toprule
\multirow{1}{*}{App} & \multirow{1}{*}{Task} & \multicolumn{3}{c}{Completion Ratio} & \multicolumn{3}{c}{E2E Execution Time (s)} \\
& & DR & CM-w/o-DR & CM & DR & CM-w/o-DR  & CM \\
\midrule
Settings & Turn on the dark theme for the system & 100\% & 100\% & 100\% & 26  & 22 & 21\\
\midrule
Chrome & Search result for today's gold price in USD in Chrome App & 73\% & 100\% & 100\% & 26 & 67 & 67 \\
\midrule
Google Play & Install RedNote App from Play Store & 33\% & 100\% & 100\% & 29 & 232 & 117 \\
\midrule
YouTube & Open the YouTube App, play the MV of Ed Sheeran's Bad Habits, and skip or close the Ad if available & 100\% & 80\% & 100\% & 66 & 600  & 88 \\
\midrule
YouTube & Open the MV of Ed Sheeran's Bad Habits on the YouTube App and open `Comments' to comment `I love this song very much' & 85\%  & 50\% & 100\% & 121 & 600 & 235 \\
\midrule
Multi-App & Search for the latest Premier League match's results in Chrome App, and write a summary in Notes & 73\% & 100\% & 100\% & 60 & 219 & 145 \\
\bottomrule
\end{tabular}
\end{table*}

\subsection{Execution-Aware Scheduling}

\method adopts an execution-aware scheduling strategy that treats backend selection as a dynamic runtime decision rather than a fixed execution path. 
Given a task, the Agent Orchestrator first consults the Memory to determine whether a structured API is available. If a matching API exists, execution proceeds through the corresponding deterministic backend, prioritizing predictable and bounded operations.

If no direct API is available, the runtime evaluates whether a semantic UI agent is required. UI agents are invoked only when task completion depends on interpreting dynamic interface state. When neither structured APIs nor semantic agents are sufficient, the system resorts to lower-level UI control actions as a fallback mechanism.

Importantly, scheduling in \method is an iterative process rather than a single-shot step. After each backend execution, the system evaluates task progress through explicit state verification. If the task is incomplete, control returns to the orchestrator for further planning and backend selection based on the current mobile states. This loop continues until the termination criteria are met.

This scheduling strategy embodies a deterministic-first policy: structured control pathways are attempted before probabilistic UI reasoning. By making escalation explicit, the runtime reduces unnecessary model invocations, limits state serialization, and improves execution predictability. More broadly, the design frames backend selection as a resource-aware scheduling problem, where reasoning cost, execution reliability, and coverage must be jointly balanced under mobile constraints.

\subsection{Implementations}

\method is implemented as an on-device runtime that integrates OpenClaw as the orchestration layer and exposes multiple control backends through a unified tool interface. The prototype runs entirely on commodity Android hardware and communicates with users through a chat channel.

The Agent Orchestrator is built on top of OpenClaw’s agent framework, which provides task interpretation, tool invocation, and iterative planning capabilities. Control Backends are implemented as modular components that expose structured action primitives. 
In our prototype, deterministic pathways are implemented via ADB-based system commands and hardware interfaces exposed through Termux API~\cite{termuxapi}. For semantic UI interaction, we employ DroidRun~\cite{droidrun} as the UI agent backend, leveraging its program-level action synthesis to support multi-step execution. These components collectively instantiate the heterogeneous control backends described in our runtime architecture.
Each backend returns bounded, machine-readable results to the orchestrator, ensuring explicit execution semantics and avoiding implicit inference of device state.
The runtime maintains a lightweight Memory component that stores execution preferences, reusable task patterns, and mobile-specific operational knowledge. This component influences backend selection and task decomposition but remains decoupled from low-level control logic. This knowledge is embedded into the context of the model or stored in the agent's retrieval path through OpenClaw's memory mechanism.

The implementation emphasizes modularity and backend extensibility. New control pathways can be integrated without modifying the orchestrator’s planning logic, enabling \method to serve as a flexible experimental platform for studying smartphone-native agent runtime design.

\section{Preliminary Results}
\subsection{Experimental Setup}
To evaluate \method, we run experiments on a Google Pixel~9 device (Android~16) and compare our \method (CM) against \emph{DroidRun} (DR), and \emph{ClawMobile without DroidRun} (CM-w/o-DR). CM-w/o-DR is a naive baseline version of \method without the advanced UI-centric mobile agent. We evaluate six real-world tasks spanning system settings, single-app operations, and cross-app workflows (Table~\ref{tab:main}). For each task, we report: (1) \textit{Completion Ratio}, a human-annotated score from 0\% to 100\% measuring task completion relative to a human (100\% indicates full completion); and (2) \textit{Time}, the end-to-end latency from issuing the instruction to the agent’s final response or completion signal (in seconds). All agents use the same underlying LLM (GPT-5.2).

\textbf{Deployment.} 
 \method and \method-w/o-DR runs directly on-device with no tethering to other devices. We connect them to a Telegram bot\footnote{https://docs.openclaw.ai/channels/telegram} so that users can submit instructions via chat, and the agent executes locally on-device. For \emph{DroidRun}, we follow the official setup guide\footnote{https://docs.droidrun.ai/v3/quickstart}: the Pixel~9 is connected via USB to a host machine with USB debugging enabled; DroidRun runs in a Python virtual environment on the host, and tasks are issued via terminal commands that drive automation on the connected device.

\subsection{Experimental Results}
\textbf{Overall Performance.}
Table~\ref{tab:main} presents the completion ratios and end‑to‑end execution times for \method, DroidRun, and ClawMobile-without-DroidRun across representative tasks. 
While \method consistently achieves near‑perfect completion (100\% on listed tasks), it incurs a higher time cost, on average 57.5 seconds slower than DroidRun. 
This trade‑off stems from a fundamental difference in the runtime strategy.
Although the CM-w/o-DR can also achieve a high completion ratio, it incurs a significantly higher time cost than \method. 
Especially on the YouTube app, it fails to complete the task due to timeouts.

\noindent\textbf{Case Study: Cross-App Search and Summarization.}
To illustrate these differences, we examine a representative multi-app task: \emph{``Search for the latest Premier League match results in Chrome, and write a summary in Notes.''} This workflow requires switching between Chrome and Notes, extracting information from the browser, and composing a short summary in a note. It reflects a common yet challenging scenario for mobile agents.

\noindent\textbf{DroidRun failures.} An UI agent fails in several ways:
\begin{enumerate}
  \item \textbf{Asynchronous app launch.} Chrome may take longer to launch than DroidRun expects, causing a false failure, such as \emph{``Precondition (Chrome open) was not met (currently on Pixel Launcher).''}
  \item \textbf{Ambiguous UI targets.} After Chrome opens, DroidRun may tap the search box but instead activate a search-history entry, leading to irrelevant results and repeated attempts to relocate the search field and correct the input; if this loop stalls, DroidRun reports failure.
  \item \textbf{Incorrect success detection.} After retrieving match results, DroidRun may open Notes but fail to create a new note due to a tapping error; in some runs, it still incorrectly concludes that the goal was achieved.
\end{enumerate}

\noindent\textbf{How \method mitigates these failures.} 
\method reduces these errors through explicit verification and recovery. 
The orchestrator treats each backend invocation (whether an ADB command or a UI-agent step) as a discrete action with a well-defined expected outcome. 
After each action, it re-queries the device state to verify progress and advances only when the intended state is reached. 
For example, in failure case (1), \method detects that Chrome is still launching and retries or waits until the browser is foregrounded, rather than terminating early. 
More generally, when a step fails, the orchestrator re-evaluates the current context, generates an updated instruction, and selects an appropriate backend controller instead of repeatedly issuing the same UI action. 
This iterative \emph{verify-and-recover} loop reduces failures that arise when control is delegated to a single backend without runtime-level checks.
\section{Challenges and Research Questions}

\method serves as a concrete instantiation of smartphone-native agentic systems, but the broader goal of this work is to articulate the system design space that governs mobile autonomy. Based on our architectural exploration and deployment experience, we group emerging challenges into three interrelated dimensions: efficiency, adaptability, and stability. Together, these dimensions define the system foundations of robust smartphone-native agents.

\subsection{Efficiency: Managing Reasoning and State as Runtime Resources}

Mobile environments impose strict constraints on computation, network bandwidth, and response time. Unlike cloud-hosted agents, smartphone-native agents must treat data transmission and LLM reasoning as costly resources.

\noindent\textbf{Efficient State Representation and Token-Aware Execution.} UI-centric agents frequently serialize large interface representations, such as full UI trees or screenshots~\cite{droidrun, wang2024mobile}, into model contexts. In mobile settings, repeated state encoding can dominate token usage and latency. Future work should explore incremental state representation, delta-based updates, and demand-driven state retrieval mechanisms that minimize unnecessary context growth while preserving task correctness. Treating the token budget as a formal runtime constraint may enable adaptive policies that explicitly trade reasoning depth for execution efficiency.

\noindent\textbf{Hybrid Deterministic–Probabilistic Scheduling.} Efficiency is tightly coupled with backend selection. Deterministic control pathways offer predictable semantics and bounded cost, while probabilistic UI reasoning provides broader coverage at higher uncertainty and token overhead. Rather than defaulting to UI-level reasoning, smartphone-native runtimes can frame backend selection as a scheduling problem under cost and reliability constraints. Formalizing hybrid execution policies, potentially incorporating confidence estimates or cost models, remains an open system question. 

\noindent\textbf{Model Placement and Runtime Partitioning.} Currently, \method runs locally, while model inference is performed remotely. Exploring on-device inference capabilities represents a promising direction~\cite{flexinfer}. Inference placement further shapes runtime efficiency. Fully remote inference introduces network latency and privacy concerns, while fully on-device models are constrained by hardware and energy budgets. Hybrid deployment strategies—such as lightweight local reasoning combined with selective remote escalation—offer a promising direction. Treating model placement as part of the runtime policy enables joint optimization of latency, cost, and autonomy in mobile agents.

\begin{tcolorbox}[colback=gray!5!white,colframe=blue!75!black,top=5pt,bottom=5pt]
\textbf{RQ1:} How can mobile agents minimize computational latency, token cost, and communication overhead while preserving task reasoning accuracy?
\end{tcolorbox}

\subsection{Adaptability: Structuring Behavior in Heterogeneous App Ecosystems}

Smartphone-native agents must operate across diverse and evolving application ecosystems. Adaptability extends beyond reasoning accuracy to structural mechanisms that support reuse, specialization, and long-term coherence.

\noindent\textbf{Structured Skill Abstraction and Domain Specialization.} Repeated low-level interaction over raw UI elements can lead to brittle execution. Introducing structured skill abstractions, i.e., reusable execution units that encapsulate common workflows, can reduce planning depth and constrain the action search space. Future research may explore automated skill extraction from execution traces, domain-specific specialization without sacrificing generality, and compositional skill design across heterogeneous apps. Such abstractions can bridge the gap between general language reasoning and app-specific operational stability.

\noindent\textbf{Persistent and Hierarchical Memory.} Mobile autonomy unfolds over time, often requiring continuity across sessions and application contexts. However, persistent memory introduces challenges related to staleness, storage efficiency, and privacy boundaries. Hierarchical memory models, which distinguish short-term execution context from longer-term summaries or skills, may offer a principled balance between adaptability and efficiency. Integrating memory management with scheduling and state abstraction remains a fertile area for runtime research.

\begin{tcolorbox}[colback=gray!5!white,colframe=blue!75!black,top=5pt,bottom=5pt]
\textbf{RQ2}: How can system and storage support agent memory adaptability across diverse and evolving mobile applications?
\end{tcolorbox}

\subsection{Stability: Reliability, Fault Tolerance, and Privacy}

Beyond efficiency and adaptability, smartphone-native agents must be reliable in deployment, balancing unstable execution conditions with strict security and privacy constraints imposed by OS and app policies.

\noindent\textbf{Reliability and Fault Tolerance.} Mobile agents often encounter partial or silent failures, e.g., interrupted control channels, actions that do not take effect, or delayed state propagation, without explicit error signals. Reliable runtimes therefore need explicit progress verification, bounded execution scopes with well-defined outcomes, and principled recovery mechanisms (e.g., handling prompts, restoring focus, relaunching, or re-navigating) to re-establish control and resume safely. Elevating reliability to a first-class runtime objective, rather than an implementation detail, is essential to move from prototypes to deployable agents.

\noindent\textbf{Security and Privacy in Mobile Autonomy.} Smartphone-native agents operate over sensitive user data and personal applications, so execution policies must incorporate privacy-aware state handling, bounded logging, and explicit permission mediation. Hybrid inference strategies further introduce trust boundaries between local and remote components, raising questions about what state is exported, under what guarantees, and how user control is preserved. Future work should formalize security and privacy models for mobile agent runtimes to ensure that autonomy does not compromise data confidentiality or user intent.

\begin{tcolorbox}[colback=gray!5!white,colframe=blue!75!black,top=5pt,bottom=5pt]
\textbf{RQ3:} How can runtime mechanisms ensure reliable, fault-tolerant, and privacy-preserving execution in unpredictable mobile environments?
\end{tcolorbox}
\section{Conclusion}

Smartphone-native agentic systems represent a new frontier in LLM-driven autonomy, where probabilistic language reasoning must operate under the constraints of real-world mobile devices. Through \method, we have explored a hierarchical runtime perspective that separates high-level reasoning from structured device control and highlighted the architectural tensions that arise in mobile environments. Our experience suggests that robust mobile autonomy cannot rely solely on improved UI reasoning or larger models, but instead requires principled coordination among control pathways, state representation, scheduling, memory, and reliability mechanisms.

By framing mobile agents as a runtime systems problem, this work shifts attention from isolated algorithmic advances toward integrated architectural design. The three dimensions discussed, efficiency, adaptability, and stability, suggest that smartphone-native agentic systems demand a unified runtime perspective. Rather than optimizing individual components in isolation, future research must co-design reasoning, control, memory, and reliability under real-device constraints. We believe this perspective lays the groundwork for advancing mobile LLM autonomy beyond UI-centric automation toward principled, smartphone-native agent architectures.

\balance
\bibliographystyle{ACM-Reference-Format}
\bibliography{refer}

\appendix

\end{document}